\documentclass[showpacs,twocolumn,aps,epsfig]{revtex4}
\usepackage{graphicx}
\begin{document}

\title{
Ab initio study of misfit dislocations at the SiC/Si(001)
interface}

\author{Giancarlo~Cicero}
\affiliation{INFM and Physics Department, Polytechnic of Torino,
C.Duca degli Abruzzi, 24,
       I-10129 Torino, Italy}
\author{Laurent~Pizzagalli}
\affiliation{LMP-CNRS, BP 30179, F-86962 Futuroscope Chasseneuil
Cedex, France}
\author{Alessandra~Catellani}
  \email{catellani@maspec.bo.cnr.it}
\affiliation{CNR-IMEM, Parco Area delle Scienze, 37A, I-43010
Parma, Italy }
\date{\today}


\begin{abstract}
The high lattice mismatched SiC/Si(001) interface was investigated
by means of combined classical and ab initio molecular dynamics.
Among the several configurations analyzed, a dislocation network
pinned at the interface was found to be the most efficient
mechanism for strain relief. A detailed description of the
dislocation core, and the related electronic properties, are
discussed for the most stable geometry: we found interface states
localized in the gap that may be source of failure of electronic
devices.
\end{abstract}


\maketitle

In recent years the development of several sophisticated epitaxial
growth techniques has allowed to produce new interesting materials
and heterostructures, even in the presence of high lattice
mismatch between the constituent phases. Furthermore,
high-resolution electron microscopy (HREM) and other similar
techniques deepened the characterization of such systems,
providing a description of the interface quality and epitaxial
relations of the sublattices. Nevertheless, the atomic structure
and the chemical environment at the interface, which deeply affect
its physical properties, still remain hardly accessible to any
experimental technique.

Atomistic simulations represent a powerful tool to complement the
experimental data. Though recent theoretical works appeared on the
study of lattice mismatched interfaces for some specific system
involving ceramic materials \cite{benedek}, the ab initio study of
a mismatched semiconductor/semiconductor interface, where the
complex covalent bonding requires an accurate description of
charge transfer \cite{ref14}, is far to be complete \cite{inas}.
Among all the possible high misfit heterostructures, we focussed
on the (001) interface between cubic silicon carbide ($\beta$-SiC)
and silicon (Si) without lacking of generality. This system,
characterized by a peculiar $\sim$20$\%$ difference in lattice
parameters between the two constituents, can be considered as a
template of high lattice mismatched heterojunctions, where an ab
initio approach is still affordable. Furthermore, it is
technologically interesting, because of the outstanding physical
properties of $\beta$-SiC, a possible candidate for applications
in harsh environment devices \cite{capano}. Some recent results on
the study of this heterostructure \cite{exp} evidenced that the
interface can be prepared so to be locally abrupt, with a network
of pure edge misfit dislocations localized at the interface to
accommodate the extreme lattice mismatch. The formation of a
dislocation network as an efficient mechanism of strain release
usually occurs in heterostructures with a mismatch larger than
$\sim$10$\%$ \cite{trampert}.

The occurrence of edge dislocations can be treated in the frame of
the well established near coincidence lattice model \cite{epi},
which asserts that the lowest interfacial energy configurations
are obtained when a perfect coincidence site between two
dissimilar structures is realized. Given two materials with
lattice parameters $\bf a_1$ and $\bf a_2$, a perfect coincidence
site occurs when $\bf a_1/a_2$=m/n, with m and n positive
integers. For the $\beta$-SiC/Si(001) system, m=5 and n=4
respectively: since m=n+1, only one extra lattice plane is needed
along each of the two primitive directions of the coincidence
lattice cell to release almost all the misfit strain, in agreement
with HREM images \cite{exp,oldapl}. A network of misfit
dislocations will thus be generated in the SiC film starting at
the interface.

In order to characterize the heterostructure and provide an
accurate description of its structural and electronic properties,
we performed a combined classical and ab initio molecular dynamics
study of different dislocation core structures: the minimum energy
system, {\sl i.e.} the most efficient way for strain release, is
obtained for dislocations pinned at the SiC/Si interface,
originating at a sub-stoichiometric, C terminated layer.

\begin{figure}
\includegraphics[width=12cm]{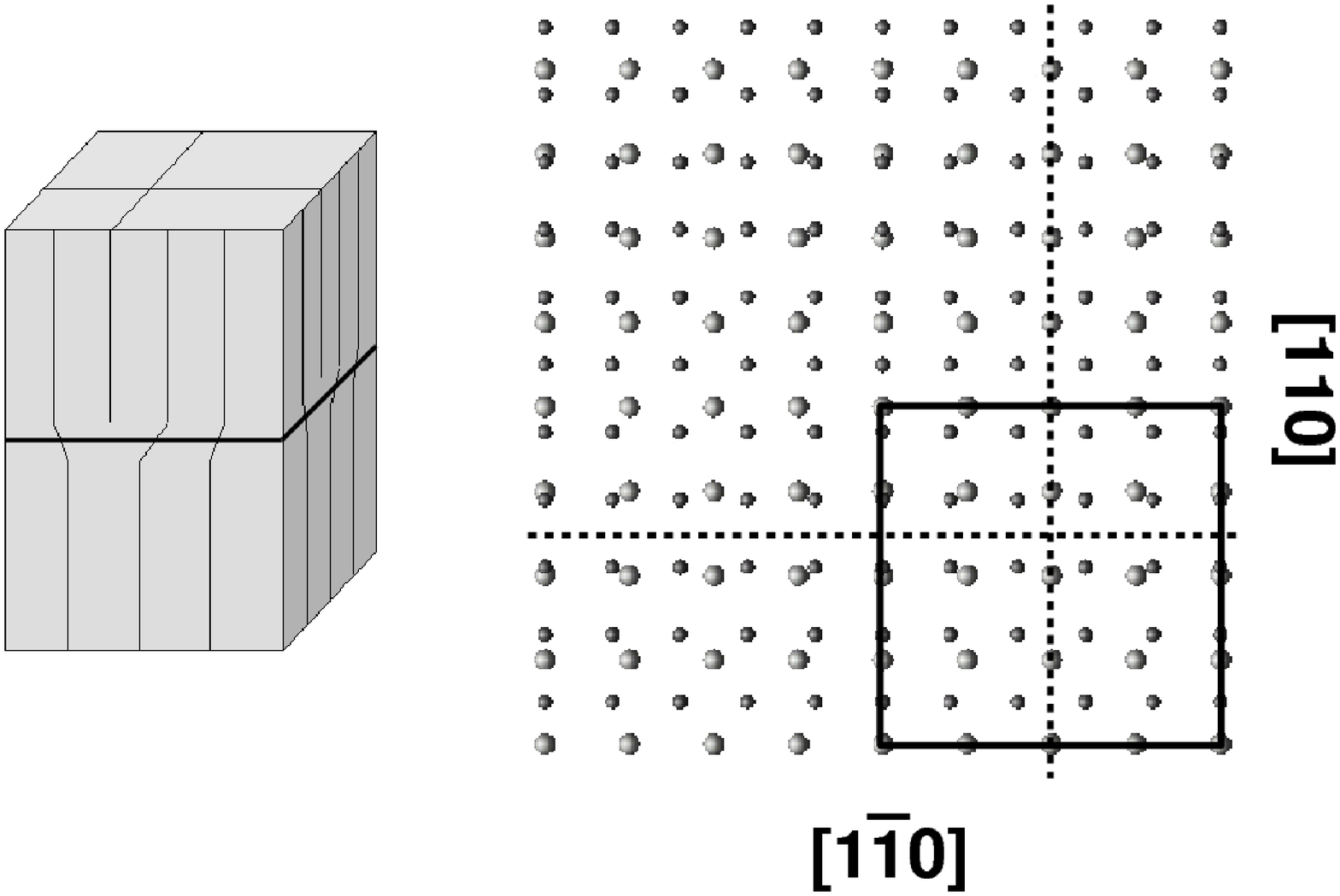}
\caption{Schematic view of the slab (left) and atomic positions of
the unrelaxed C-Si interface projected along (001) (right). The
computational unit cell (cubic coincidence lattice) is indicated
in continuous lines and the edge dislocation lines [110] and
[1\={1}0] are evidenced. Black (grey) spheres represent C (Si)
atoms. Only two planes at the interface are shown.}\label{fig1}
\end{figure}

To reproduce the experimental situation, we considered a
periodically repeated multilayer: the interface was built by
matching a p(4x4)-Si(001) with a p(5x5)-SiC(001) slab, at the Si
lattice parameter optimized for bulk calculations with the chosen
potentials (either classical or ab-initio). This model contains
two edge dislocations in the supercell, one along each of the two
directions perpendicular to the interface, and the periodicity
experimentally observed is thus respected. In Fig.~\ref{fig1} the
interface structure is schematically represented and the
coincidence lattice supercell, which we considered in our
calculations, is evidenced. At the interface, the dislocation
directions are not equivalent, because of the zincblende stacking:
this renders inequivalent  the dislocation cores.

We have performed a structural optimization for several possible
interface configurations, varying both geometry and stoichiometry,
via classical and ab initio Molecular Dynamics (MD) methods in
order to get the most stable structure. The relative energy of
systems containing different numbers of atoms has been discussed
in the grand canonical scheme \cite{chempot}.

Classical MD for several interface configurations allowed us to
select the core dislocation structures and slab size for the ab
initio simulations. The Si-Si, Si-C and C-C interactions were
modelled by means of the empirical Tersoff's potentials
\cite{tersoff2}. It has recently been demonstrated that these
potentials give formation energies and properties of native
defects in $\beta$-SiC(001) in good agreement with ab initio
calculations \cite{nuclear}. We performed tests varying the
stoichiometry at the interface and the core dislocation distance
from the interface. In order to prevent the interaction between
the interface and the slab surfaces, we considered thick slabs (36
SiC layers on top of 36 Si layers: in the following, we will
indicate the systems as N/M, where N (M) stems for the number of
SiC (Si) layers, thus the slab described above will be labelled as
36/36). All the atoms were allowed to move, except those belonging
to the outermost layers, which were constrained to relax in a
collective way. Our results indicate that configurations with C
atoms at the interface are always more stable. A pseudomorphic SiC
layer on top of Si, before the dislocations start, is unfavorable.
In agreement with experimental results \cite{exp}, the favored
core structures are those with the dislocations pinned at the
interface \cite{nota}. Among all the considered geometries, we
focussed on the most meaningful ones: the stoichiometric interface
and the one which is lowest in energy. The latter is obtained from
the stoichiometric solution by removal of all the C atoms along
the two core dislocation lines (13 atoms per supercell): this
configuration gives an energy gain $\Delta$E of 1.12 J/m$^2$
(16.43 eV/cell-area) and 0.48 J/m$^2$ (7.07 eV/cell-area) for
C-poor and C-rich conditions respectively \cite{chempot}, in
classical MD and for a 36/36 slab.

In order to determine the minimum size required to get accurate
results via ab initio techniques, we analyzed how the relative
energies and atomic structures of these two systems vary as a
function of the slab thickness, in classical MD calculations. We
observed that the relative atomic distortion is less than 2 $\%$
after the 4th layer, when considering a 5/5 or a 36/36 layer slab;
furthermore, the value obtained for $\Delta$E for the two
configurations described above in a 7/7 slab reproduces correctly
the asymptotic value of an infinite (36/36) system.  When a 5/5
slab is considered, $\Delta$E decreases by $\simeq 20 \%$ in
C-poor conditions ({\sl i.e.} $\Delta$E = 12.75 eV/cell-area), due
to a stronger surface-interface interaction obtained in the
non-stoichiometric structure. These results, that provide a good
estimate for the system stability,  allowed us to consider only
the 5/5 slab in ab initio calculations.

The ab initio MD \cite{hamann} simulations were performed in the
frame of density functional theory (DFT), in the local density
approximation (LDA) \cite{LDA}. The electronic wavefunctions
(charge density) were expanded in plane waves with an energy
cutoff of 40 (160) Ry; only the $\Gamma$ point was included for
the integration in the supercell Brillouin Zone (BZ). The
electron-ion interaction was described by fully non local
pseudopotentials \cite{hamann}. Surface atoms at both sides were
saturated with hydrogen, at a distance optimized via preliminary
surface calculations \cite{nota1}. All the atoms of the slab were
allowed to move except the hydrogens and the Si (C) in the
outermost layers, which were relaxed along the direction
perpendicular to the surface only. In a subsequent test, this
constrain has been released in order to evaluate the interaction
between surfaces and interfaces.
Structures were considered converged when forces acting on atoms
were less than 10$^{-4}$ a.u. (0.005 eV/\AA) and energy varied by
10$^{-5}$ eV/atom.


We observed that the classical simulations give good initial guess
geometries for ab initio calculations, though lacking of accuracy
in predicting structures where dangling bonds are present. In
particular, at variance with the empirical simulation, the
stoichiometric system reconstructs around the two dislocation
cores, where dangling bonds are localized, if relaxed by the ab
initio technique. The core structures are thus characterized by 5-
and 7-membered rings with C-C and Si-Si dimers at the SiC and Si
interface layers respectively \cite{nota2}. The same energy
ordering was however found in the two different methods: the
sub-stoichiometric solution is the most stable in the full
physical range of variability of the C chemical potential. The ab
initio energy gain for this geometry is $\Delta$E = 14.9 (5.6)
eV/cell-area, in C-poor (C-rich) conditions.

\begin{figure}
\includegraphics[width=12cm]{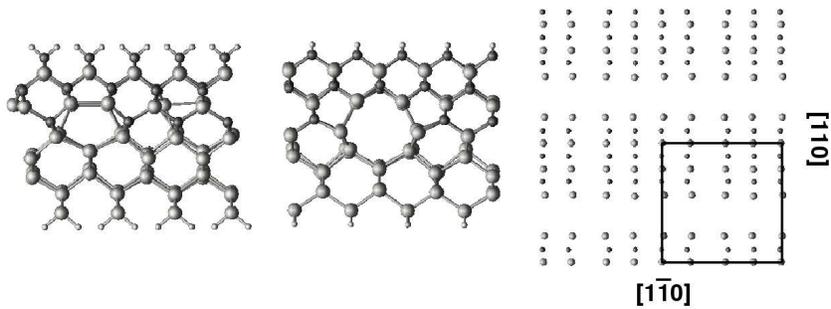}
\caption{(110) (left) and (1\={1}0) (center) projection of the
most stable interface structure. A projected view of the interface
atoms is presented as an insert (right) to enhance the warping.
Black (grey) spheres represent C (Si) species. Surface atoms are
saturated with hydrogen (small white spheres).} \label{fig2}
\end{figure}

The relaxed structure for the lowest energy interface is reported
in Fig.~\ref{fig2}. The interplanar distance at the interface is
lower (-13$\%$) than the one obtained in bulk SiC: the deformation
optimizes the SiC cell volume and bond distances at this high
lattice mismatch. We observed that both Si-Si (in Si) and Si-C
bonds (in SiC and at the SiC/Si interface) are stretched at the
dislocation cores, compared to the bulk bond lengths, being
elongated up to 2.5 {\AA} and 2.0 {\AA}, respectively. The
tetrahedral configuration is distorted too, with some Si-C-Si bond
angles close to 120$^{\circ}$. The bulk structure is quickly
recovered when moving away from the interface and at the third
atomic layer the residual distortions are already small
($\pm$3$\%$ in bond distances, interplanar spacing and angles).
The deformation induced by the interface and the dislocation cores
has been also analyzed in terms of warping of the atomic layers
(see right panel in Fig.~\ref{fig2}). We found that the major
deformations are localized in Si, since it has smaller elastic
constants than SiC. The layer puckering decreases when moving
aside the interface: this is in nice agreement with some recent
experimental results \cite{warp} on the structural
characterization of SiC films grown on a Si(001) substrate, which
evidenced an internal roughness of individual SiC planes fading
away from the Si substrate. As noticed in Fig.~\ref{fig2}, the
sub-stoichiometric structure does not present any dangling bond:
the Si atoms in the second SiC layer, which would have an
unsaturated bond, reconstruct and form dimers 2.4-2.5~{\AA} long;
this is also evidenced by the analysis of the electronic structure
(see below) which proves the presence of bonding charge density
localized between interfacial Si atoms. The removal of C atoms and
stretched C dimers in the core allows to better accommodate the
misfit strain, although maintaining an abrupt SiC/Si interface:
this stabilizes the system with respect to the stoichiometric one.

In order to have a qualitative indication of the interface/surface
interaction, we performed further calculations on the 5/5 slab,
relaxing the surface atoms at both sides. The results of these
simulations indicate that already for the 5/5 system the interface
is well described, and the interaction with the surface is small:
the bond lengths around the dislocation cores change less than
3$\%$ and the energy difference between the stoichiometric and
sub-stoichiometric slabs is modified by 18$\%$ for C-poor
conditions. The major distortions obviously occurs at surface
layers, that bent to accommodate the residual elastic strain
($\Delta$z = 0.9{\AA}).

\begin{figure}
\includegraphics[width=12cm]{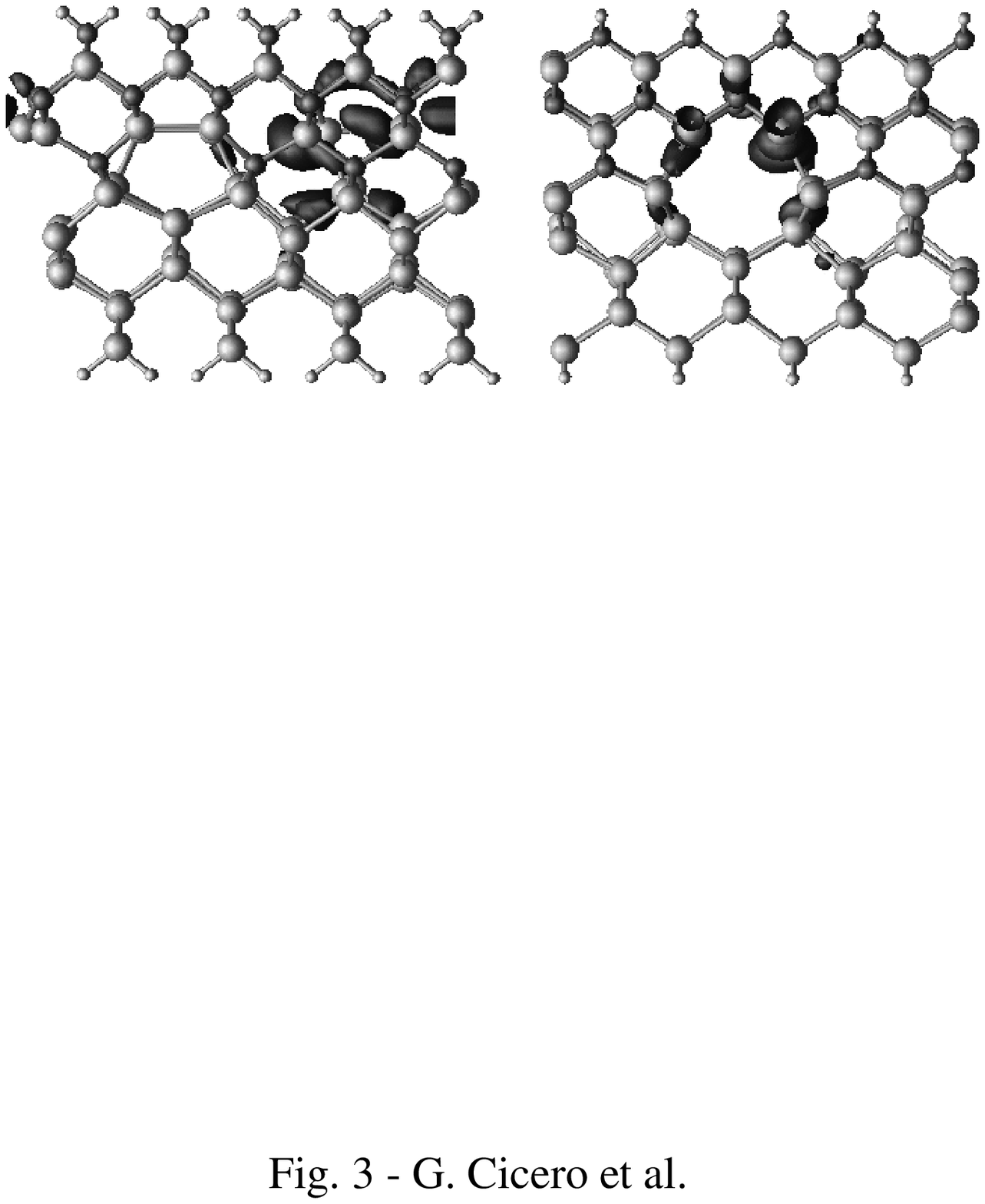}
\caption{Charge density contour plot of the highest occupied state
projected along (110) (left) and (1\={1}0) (right). } \label{fig3}
\end{figure}

The interface structure determines the electronic properties of
the heterojunction. In particular, the presence of defects such as
misfit dislocations can induce interface states, also called
interface trapped charges, localized in the band gap, that may be
responsible for device failure. We have analyzed the electronic
structure for the most stable dislocation network at the
$\beta$-SiC/Si(001) interface. Although the number of dangling
bonds is minimized in this system, the Highest Occupied (HO) and
Lowest Unoccupied (LU) states are localized in the core of the
edge dislocations, as a result of the large difference in charge
transfer between Si-C and Si-Si bonds. These states lay in the
interface band-gap, at ~0.8 and ~1.3 eV above the valence band top
at $\Gamma$ \cite{nota3}, thus they may be a perturbation to the
electron mobility and optical properties of the SiC/Si
heterostructure. In Fig.~\ref{fig3}, the charge density plot of
the HO state along the two dislocation directions is represented.
The charge density is mainly localized on atoms of the
[1$\bar{1}$0] core dislocation, while no density is observed
around the core in the perpendicular direction. The opposite
situation is found in other bonding states localized in the system
forbidden gap and for the LU state (not shown) which charge
densities pertain to the dislocation core existing along the [110]
direction.


In conclusions, using classical and first-principles calculations,
we have presented a comparative analysis of the structural and
electronic properties of edge dislocation networks at a high
lattice mismatched semiconductor/semiconductor interface. For the
studied case, $\beta$-SiC/Si(001), the core dislocations which
were found as most stable, and which are thus responsible for
strain relief in the heterostructure, are pinned at the interface,
and occur at sub-stoichiometric C layers. The presence of these
extended defects is responsible for occupied and unoccupied
states, laying in the forbidden gap, which can severely affect the
electronic properties of the system.

One of us (LP) acknowledges INFM support for his stay in Italy as
visiting professor. GC acknowledges Demichelis Foundation for his
PhD fellowship. We also wish to thank Alexis Baratoff for fruitful
discussions. Calculations performed in CINECA (IT) through the
INFM Parallel Computing Initiative and at CSCS (Manno, CH). This
work was partially supported by INFM-PRA:1MESS.

\newpage

\begin{thebibliography}{00}
%
\bibitem{benedek}
R.~Benedek, {\sl et al.},
Phys.~Rev.~Lett. {\bf 84}, 3362
(2000);
%
R.~Benedek, {\sl et al.},
Phys.~Rev.~B {\bf 60}, 16094 (1999).
%
\bibitem{ref14}
G.~Galli, F.~Gygi, and A.~Catellani, Phys. Rev. Lett. {\bf 82},
3476 (1999).
%
\bibitem{inas}
N.~Oyama, {\sl et al.},
Surf. Sci. {\bf 433-435}, 900 (1999).
%
\bibitem{capano}
See, e.g.,
M. A. Capano, and R. J. Trew, Mater. Res. Bull. {\bf 22}, 19 (1997).
%
\bibitem{exp}
C.~Long, S.~A.~Ustin, and W.~Ho,
J.~Appl.~Phys. {\bf 86}, 2509 (1999).
%
\bibitem{oldapl}
S.~R.~Nutt, {\sl et al.},
Appl.~Phys.~Lett. {\bf 50}, 203 (1987).
%
\bibitem{trampert}
A.~Trampert, and K.~H.~Ploog,
Cryst.~Res.~Technol. {\bf 35}, 793 (2000).
%
\bibitem{epi}
N.~H.~Fletcher, and K.~W.~Lodge, {\it Epitaxial Growth, Part B},
in J. W. Matthews (ed.), Academic Press, New York, 529 (1975).
%
\bibitem{chempot}
G.~X.~Qian, R.~M.~Martin and D.~J.~Chadi, Phys.~Rev.~B {\bf 38},
7649 (1988). We varied the chemical potential between the total
energy of bulk Si (as obtained consistently) and that of bulk Si
minus the heat of formation of the SiC crystal (the experimental
value is $\Delta$Hf = 0.72 eV). The equation $\mu_{SiC}$ = $\mu_C$
+ $\mu_{Si}$ holds in the full physical range: this defines C-poor
and C-rich conditions respectively as the two extrema mentioned
above.
%
\bibitem{tersoff2}
J.~Tersoff, Phys.~Rev.~B {\bf 39}, 5566 (1989).
%
\bibitem{nuclear}
F.~Gao, {\sl et al.},
Nucl.~Instr.~and~Meth.~B {\bf 180}, 286
(2001).
%
\bibitem{nota}
This corresponds to a vanishing critical thickness for SiC
overlayers on Si.
%
\bibitem{hamann}
We used the first principles molecular dynamics programs {\em
BASIC96} and {\em JEEP} (G. Galli and F. Gygi). For a review see,
e.g., G. Galli and A. Pasquarello, in {\sl Computer Simulation in
Chemical Physics}, Edited by M.P.Allen and D.J.Tildesley,  p. 261,
Kluwer, Dordrecht  (1993); and M.~C.~Payne, {\sl et
al.},
Rev. Mod. Phys. {\bf 64}, 1045  (1993).
Pseudopotentials generated as in D.R. Hamann, Phys.~Rev.~B {\bf
40},  2980  (1989). (We used {\sl s} and {\sl p} nonlocality for
Si  and {\sl s} nonlocality for C).
%
\bibitem{LDA}
W. Kohn, and L. Sham, Phys. Rev. A {\bf 140}, 1133 (1965).
%
\bibitem{nota1}
These tests were performed with symmetric slabs of 13 layers, plus
a H terminating overlayer at both surfaces. In this case, a
$p(2\times2)$ surface periodicity was considered (4 atoms/layer)
and all the atoms were allowed to move.
%
\bibitem{nota2} A full description of the energetics, the structural and
electronic properties of the investigated systems, along with a
detailed analysis of the strain field within elasticity theory
will be presented in a forthcoming publication.
%
\bibitem{warp}
G.~Xu, and Z.~C.~Feng,
Phys.~Rev.~Lett. {\bf 84}, 1926 (2000).
%
\bibitem{nota3}
The valence band widths of Si- and SiC-derived bulk states compare
fairly well for relaxed and unrelaxed surfaces, although
underestimate the respective bulk calculations. Furthermore they
change by $\simeq 4\%$ when increasing the slab thickness to 7/7.
We were thus able to estimate an error of $\pm$ 0.2 eV on LDA
eigenvalues. The energy positions of HO and LU states, with
respect to the valence band top of the bulk constituents, as
obtained in this work, are beyond these  limits.
%
\end{thebibliography}
\end{document}